\documentclass[preprint,showpacs,superscriptaddress,preprintnumbers,amsmath,amssymb]{elsart}

\usepackage{graphicx}
\usepackage{rotating}
\usepackage{dcolumn}
\usepackage{bm}

\begin{document}
\begin{frontmatter}
\title{Study of \boldmath{$\Lambda^+_c$} Cabibbo Favored 
Decays Containing a \boldmath{$\Lambda$} Baryon
in the Final State}

\date{\today}

The FOCUS Collaboration\footnote{See http://www-focus.fnal.gov/authors.html for
additional author information.}
\author[ucd]{J.~M.~Link}
\author[ucd]{P.~M.~Yager}
\author[cbpf]{J.~C.~Anjos}
\author[cbpf]{I.~Bediaga}
\author[cbpf]{C.~Castromonte}
\author[cbpf]{A.~A.~Machado}
\author[cbpf]{J.~Magnin}
\author[cbpf]{A.~Massafferri}
\author[cbpf]{J.~M.~de~Miranda}
\author[cbpf]{I.~M.~Pepe}
% ...

\address[ucd]{University of California, Davis, CA 95616}
\address[cbpf]{Centro Brasileiro de Pesquisas F\'\i sicas, Rio de Janeiro, RJ, Brazil}
\address[cinv]{CINVESTAV, 07000 M\'exico City, DF, Mexico}
\address[cu]{University of Colorado, Boulder, CO 80309}
\nopagebreak
\address[fnal]{Fermi National Accelerator Laboratory, Batavia, IL 60510}
\address[fras]{Laboratori Nazionali di Frascati dell'INFN, Frascati, Italy I-00044}
\address[ugj]{University of Guanajuato, 37150 Leon, Guanajuato, Mexico}
\address[ui]{University of Illinois, Urbana-Champaign, IL 61801}
\address[iu]{Indiana University, Bloomington, IN 47405}
\address[korea]{Korea University, Seoul, Korea 136-701}
\address[kp]{Kyungpook National University, Taegu, Korea 702-701}
\address[milan]{INFN and University of Milano, Milano, Italy}
\address[nc]{University of North Carolina, Asheville, NC 28804}
\address[pavia]{Dipartimento di Fisica Nucleare e Teorica and INFN, Pavia, Italy}
\address[po]{Pontif\'\i cia Universidade Cat\'olica, Rio de Janeiro, RJ, Brazil}
\address[pr]{University of Puerto Rico, Mayaguez, PR 00681}
\address[sc]{University of South Carolina, Columbia, SC 29208}
\address[ut]{University of Tennessee, Knoxville, TN 37996}
\address[vu]{Vanderbilt University, Nashville, TN 37235}
\address[wisc]{University of Wisconsin, Madison, WI 53706}

\begin{abstract}

Using data from the FOCUS experiment (FNAL-E831), we study  the decay of
$\Lambda^+_c$ baryons into final states containing a
$\Lambda$ hyperon.
The branching fractions of $\Lambda^+_c$ into $\Lambda \pi^+$, $\Lambda \pi^+ \pi^+ \pi^-$ and 
$\Lambda \overline{K}\,\!^0 K^+$ relative to that into $pK^-\pi^+$ are measured to
be $0.217 \pm 0.013  \pm 0.020$, $0.508 \pm 0.024  \pm 0.024$ and 
$0.142 \pm 0.018 \pm 0.022$, respectively. 
We also report new measurements of  
$\frac{\Gamma(\Lambda^+_c \rightarrow \Sigma^0 \pi^+)}{\Gamma(\Lambda^+_c \rightarrow
\Lambda \pi^+)} = 1.09 \pm 0.11 \pm 0.19$, 
$\frac{\Gamma(\Lambda^+_c \rightarrow \Sigma^0 \pi^+\pi^+ \pi^-)}{
\Gamma(\Lambda^+_c \rightarrow \Lambda \pi^+ \pi^+ \pi^-)} = 
0.26 \pm 0.06 \pm 0.09$ and $\frac{\Gamma(\Lambda^+_c \rightarrow 
\Xi(1690)^0(\Lambda \overline{K}\,\!^0) K^+)}{
\Gamma(\Lambda^+_c \rightarrow \Lambda \overline{K}\,\!^0 K^+)} = 0.33 \pm 0.10
\pm 0.04$.
Further, an analysis  
of the subresonant structure for the $\Lambda^+_c \rightarrow 
\Lambda \pi^+\pi^+\pi^-$ decay mode is presented. 
\end{abstract}
\end{frontmatter}

\section{\label{sec:level1}Introduction}

During the past several years there has been significant progress in the experimental study of
hadronic decays of charmed baryons. However the precision on branching fraction measurements
is only about 40\% for many Cabibbo-favored modes and even worse for
Cabibbo-suppressed decays~\cite{PDG}. 
As a result, we are not yet able to distinguish between the decay
rate predictions made by different theoretical models, e.g., the quark model approach to
non-leptonic charm decays and the Heavy Quark Effective Theory (HQET)~\cite{theo1,theo2,theo3}.
In this paper we present a study of $\Lambda^+_c$ baryons produced by the FOCUS experiment. We
present improved measurements of the branching fractions of the 
Cabibbo-favored decays $\Lambda^+_c \rightarrow \Lambda 
\pi^+, \Lambda^+_c \rightarrow \Lambda \pi^+\pi^+\pi^-$ and $
\Lambda^+_c \rightarrow \Lambda \overline{K}\,\!^0 K^+$. From the measurement of the first 
two modes, we are also able to extract the relative branching ratios of the two decays  
$\Lambda^+_c \rightarrow \Sigma^0 \pi^+$ and $\Lambda^+_c \rightarrow \Sigma^0
\pi^+\pi^+\pi^-$. We report a new measurement of the subresonant mode $\Lambda^+_c \rightarrow
\Xi(1690)^0 K^+$.
Finally we present the first study of the subresonant structure of the 
$\Lambda^+_c \rightarrow \Lambda \pi^+\pi^+\pi^-$ decay mode.

\section{\label{sec:level2}Event Reconstruction}
This analysis uses data collected by the FOCUS experiment during the 1996--1997
fixed-target run at Fermilab. 
  
FOCUS is a photo-production experiment equipped with very precise vertexing and
particle identification detectors. The vertexing system is composed of a silicon
microstrip detector (TS) embedded in the BeO target segments~\cite{TSNIM}, and a
second system of twelve microstrip planes (SSD) downstream of the target. Downstream of
the SSD, five stations of multiwire proportional chambers and two large aperture
dipole magnets complete the charged particle tracking and momentum measurement system.
Three multicell threshold \v{C}erenkov detectors are used to identify 
electrons, pions, kaons, and protons.
The FOCUS apparatus also contains one hadronic and two electromagnetic
calorimeters as well as two muon detectors. 
 

All decay modes reported have a $\Lambda$ hyperon\footnote{Throughout this paper
the charged conjugate state is implied unless explicitly stated.}
in the final state. A detailed
description of $\Lambda$ and $K^0_S$ reconstruction techniques in FOCUS is
reported in Reference~\cite{vees.0109028}.
  
Candidates are reconstructed by first forming a vertex with tracks consistent
with a specific $\Lambda^+_c$ decay hypothesis. A cut on the confidence level that these
tracks form a good vertex is applied. Production vertex candidates are found using
a candidate driven vertexing algorithm which uses the 
$\Lambda^+_c$ candidate momentum to define the line of the flight of the charm
particle~\cite{nim.320.519}. This seed track is intersected with other
tracks in the event to form a production vertex. 
The confidence level for the production vertex must be greater than 1\%.
Most of the background is rejected by applying a separation cut between the
production and decay vertices: we require the significance of separation between
the two vertices, $L/\sigma_L$, to be greater than some number, depending on the
decay mode.

All charged microstrip track segments from the charm decay must be linked to a single multi-wire proportional chamber 
track segment, be of good quality,
and be inconsistent with zero degree tracks from photon conversion. 
The likelihood for each charged particle to be proton, kaon, pion or electron
based on \v{C}erenkov particle identification is used to make additional
requirements~\cite{ceren.0108011}. 
For pion candidates, we require a loose cut that no
alternative hypothesis is favored over the pion hypothesis by more than 6 units
of log-likelihood. In addition, for each kaon candidate we require the negative
 log-likelihood kaon hypothesis, $W_K = -2 \ln $ (kaon likelihood), to be
 favored over the corresponding pion hypothesis $W_{\pi}$ by $W_{\pi} -W_K > 3$.

The reconstructed mass of the $\Lambda$ candidates must be between 1.1 and 1.125
GeV/$c^2$; no cut is applied on the normalized mass $[M(\Lambda)
-M(\Lambda)_{PDG}]/\sigma_{M(\Lambda)}
$, because it is not centered
 around zero, probably due to the higher background under the signal region. 
 We moreover require the higher momentum track used to reconstruct the 
 $\Lambda$ candidates to be compatible with the proton hypothesis, 
 applying the cut $W_{\pi} -W_p > 4$.
The reconstructed mass of the $K^0_S$ must be within three standard 
deviations of the nominal $K^0_S$ mass.

We require $\Lambda^+_c$ candidates to have a minimum momentum of 45 GeV/$c$ and
to have a lifetime less than five times the nominal value~\cite{PDG}. 
 Finally, in order to reduce backgrounds, we require the production vertex to be
 located inside the target material. 
\vspace{-0.31cm}
\section{\label{sec:level3}The normalization mode}

The $\Lambda^+_c \rightarrow p K^- \pi^+$ channel is our highest statistics
$\Lambda^+_c$ decay mode and it is used as the normalization
mode for branching ratio measurements to minimize the overall statistical 
uncertainty. Moreover, all previous measurements in the literature~\cite{PDG} 
use this decay as a normalization mode, 
thus making any comparison straightforward. 

In order to minimize systematic biases, the normalization mode is selected using
the same cuts and the same fitting technique as the specific decay 
whenever possible. In addition, for each proton candidate we apply the cuts $W_{\pi} -W_p > 4$
and $W_K -W_p > 1$.  
The $pK^-\pi^+$ invariant mass distribution for an $L/\sigma_L > 4$ cut is shown in 
Fig. 1 (b). The resultant yield is $16447 \pm 193 $ events.

\section{\label{sec:level4} The $\Lambda^+_c \rightarrow \Lambda \pi^+$ 
decay mode}

We measure the branching ratio of $\Lambda^+_c \rightarrow \Lambda \pi^+$
relative to $\Lambda^+_c \rightarrow p K^- \pi^+$. In Fig. 1 (a)
the $\Lambda \pi^+$
invariant mass distribution for an $L/\sigma_L > 4$ cut is presented.
The confidence level for the decay vertex must be greater than 1\%. 
We also apply a $|\cos \theta| <$ 0.6  cut, where $\theta$
is the angle between the $\Lambda$ momentum in the $\Lambda^+_c$ rest frame 
and the $\Lambda^+_c$ laboratory momentum.

\begin{figure}[!tbp]
     \includegraphics[width=6.855cm,height=6.89cm]{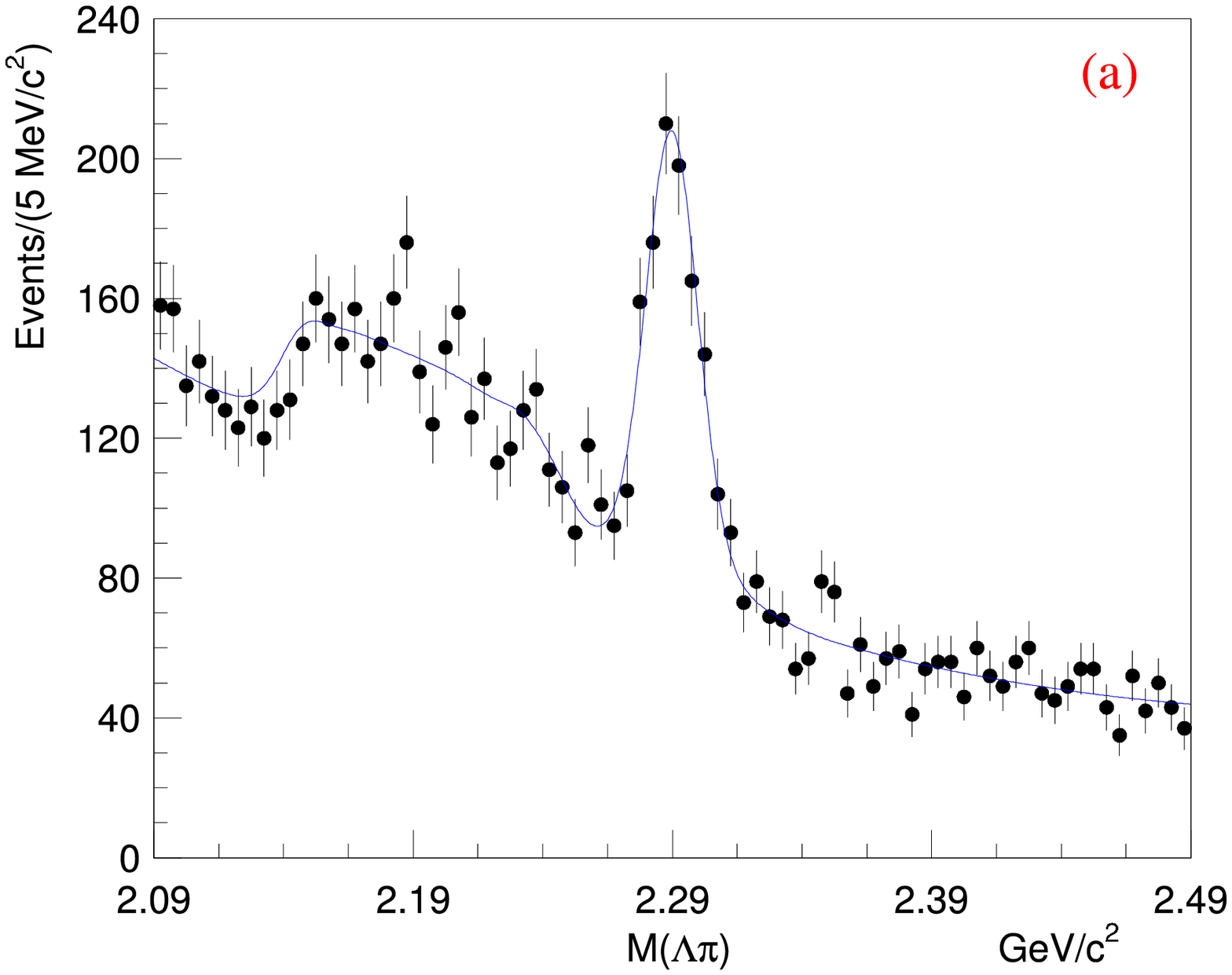}
     \includegraphics[width=6.825cm,height=6.8cm]{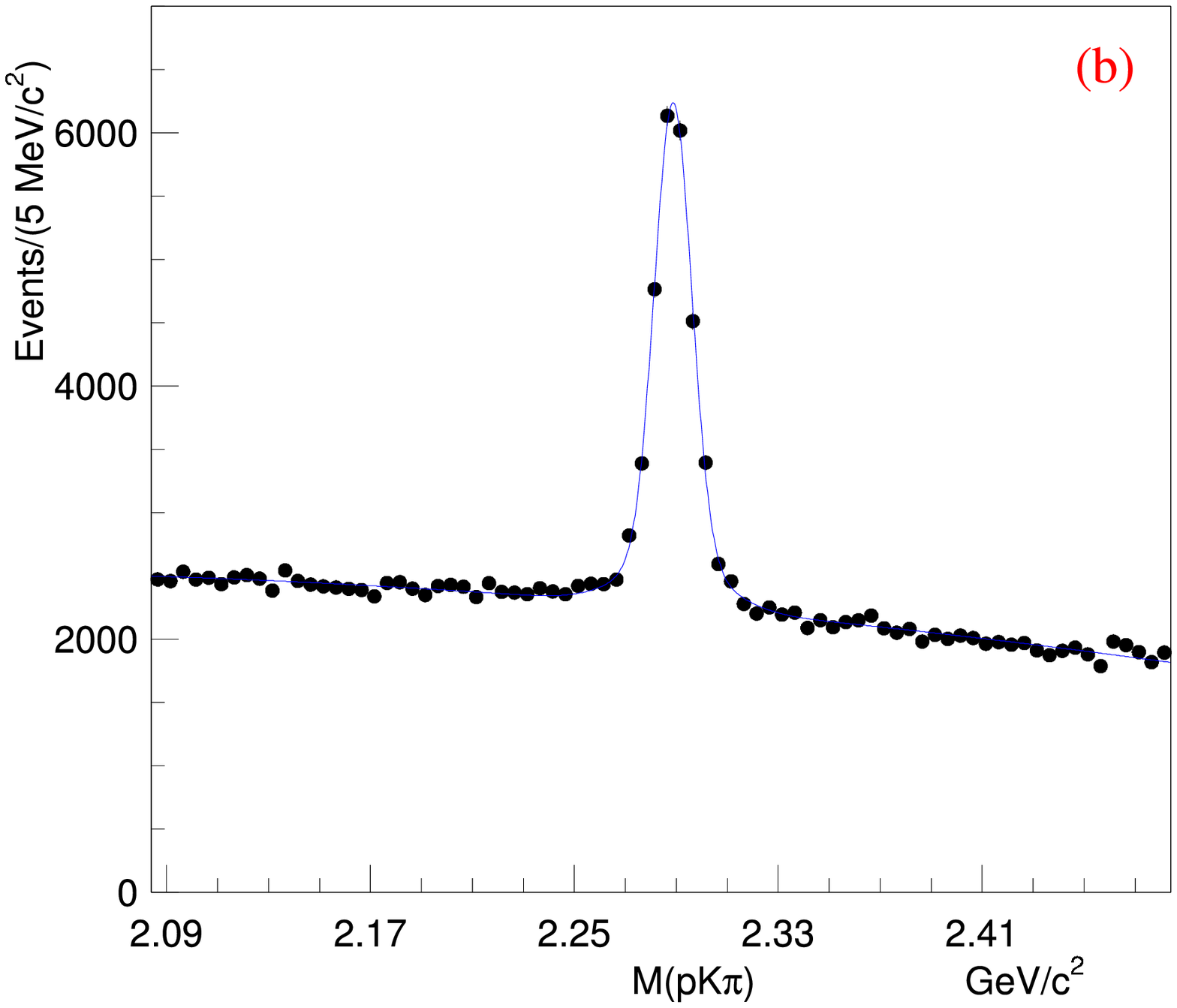}
     \vspace{-0.7cm}
  \label{figuno:mass}
  \caption{Invariant mass distributions for (a) $\Lambda^+_c \rightarrow 
  \Lambda \pi^+$ and (b) $\Lambda^+_c \rightarrow p K^- \pi^+$. The fits are
  described in the text. }
\end{figure}

We note a broad structure around 2.2 GeV/$c^2$ coming from the decay mode 
$\Lambda^+_c \rightarrow \Sigma^0 (\Lambda \gamma) \pi^+$ where the photon from
the $\Sigma^0$ decay is not reconstructed. The shape for
this reflection has been obtained from a Monte Carlo simulation of this decay
mode. The fit is performed using two Gaussians with the same mean for the 
signal, the reflection from the $\Sigma^0 \pi^+$ mode, and a second order 
Chebychev polynomial for the background. 
The ratio of yields and the resolutions of the two Gaussians are fixed to the 
Monte Carlo values.
The resultant yield is $750 \pm 44 $ events. Correcting for the relative
efficiencies estimated by our Monte Carlo simulation, 
we determine the branching ratio to be

\begin{equation} 
\frac{\Gamma(\Lambda^+_c \rightarrow \Lambda \pi^+)}
{\Gamma(\Lambda^+_c \rightarrow p K^-\pi^+)}= 0.217 \pm 0.013~(\mbox{stat.}). 
\end{equation} 

 The number of fitted $\Lambda^+_c \rightarrow \Sigma^0 \pi^+$ reflection events
is 919 $\pm$ 92. Correcting for the relative efficiencies, 
we extract the relative branching ratio:

\begin{equation}
\frac{\Gamma (\Lambda^+_c \rightarrow \Sigma^0 \pi^+)}
{\Gamma (\Lambda^+_c \rightarrow \Lambda \pi^+)} = 1.09 \pm 0.11~(\mbox{stat.}).
\end{equation}

\section{\label{sec:level5} The $\Lambda^+_c \rightarrow \Lambda \pi^+ \pi^+
\pi^-$ decay mode}

\begin{figure}[!tbp]
     \includegraphics[width=6.825cm,height=6.8cm]{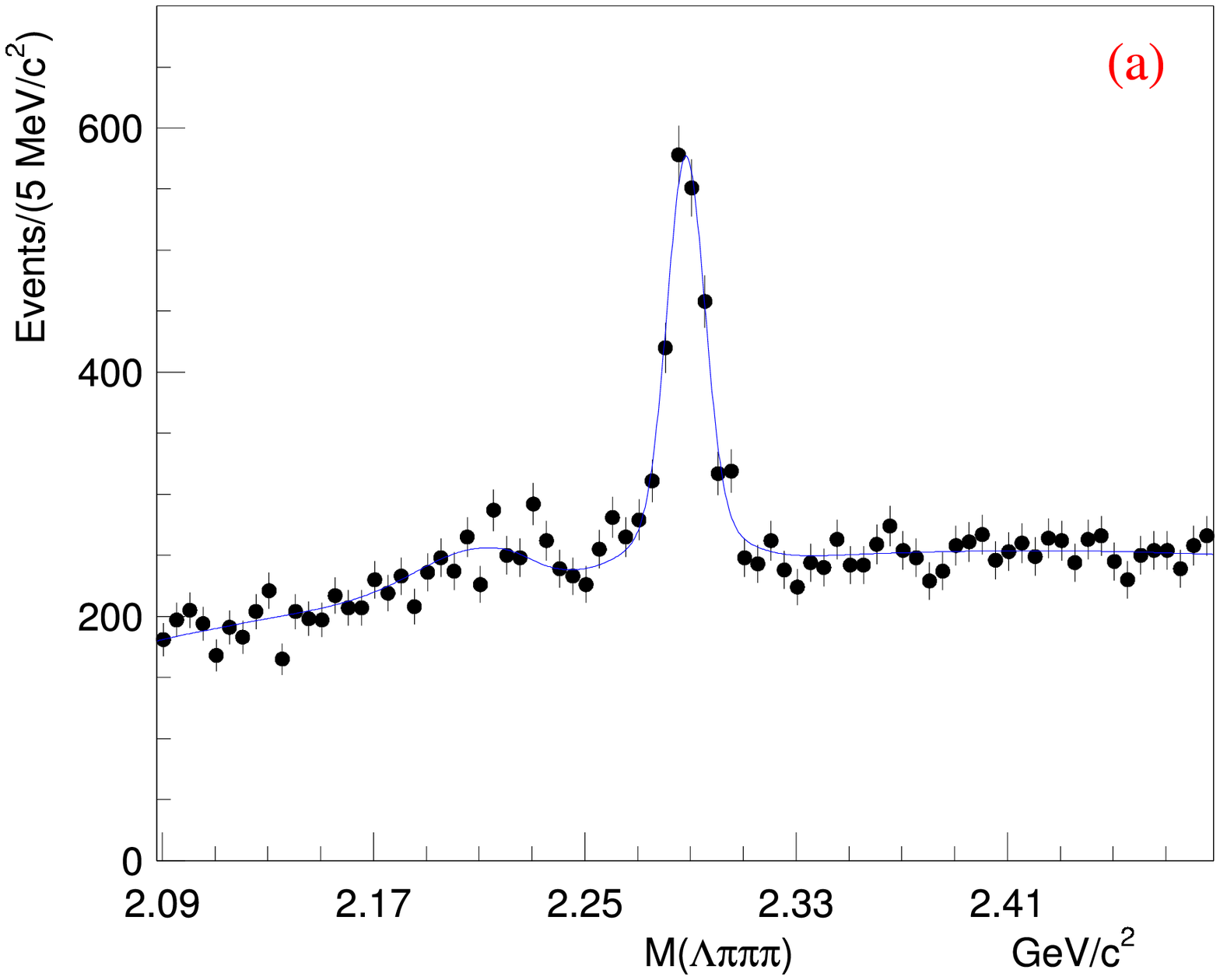}
     \includegraphics[width=6.825cm,height=6.89cm]{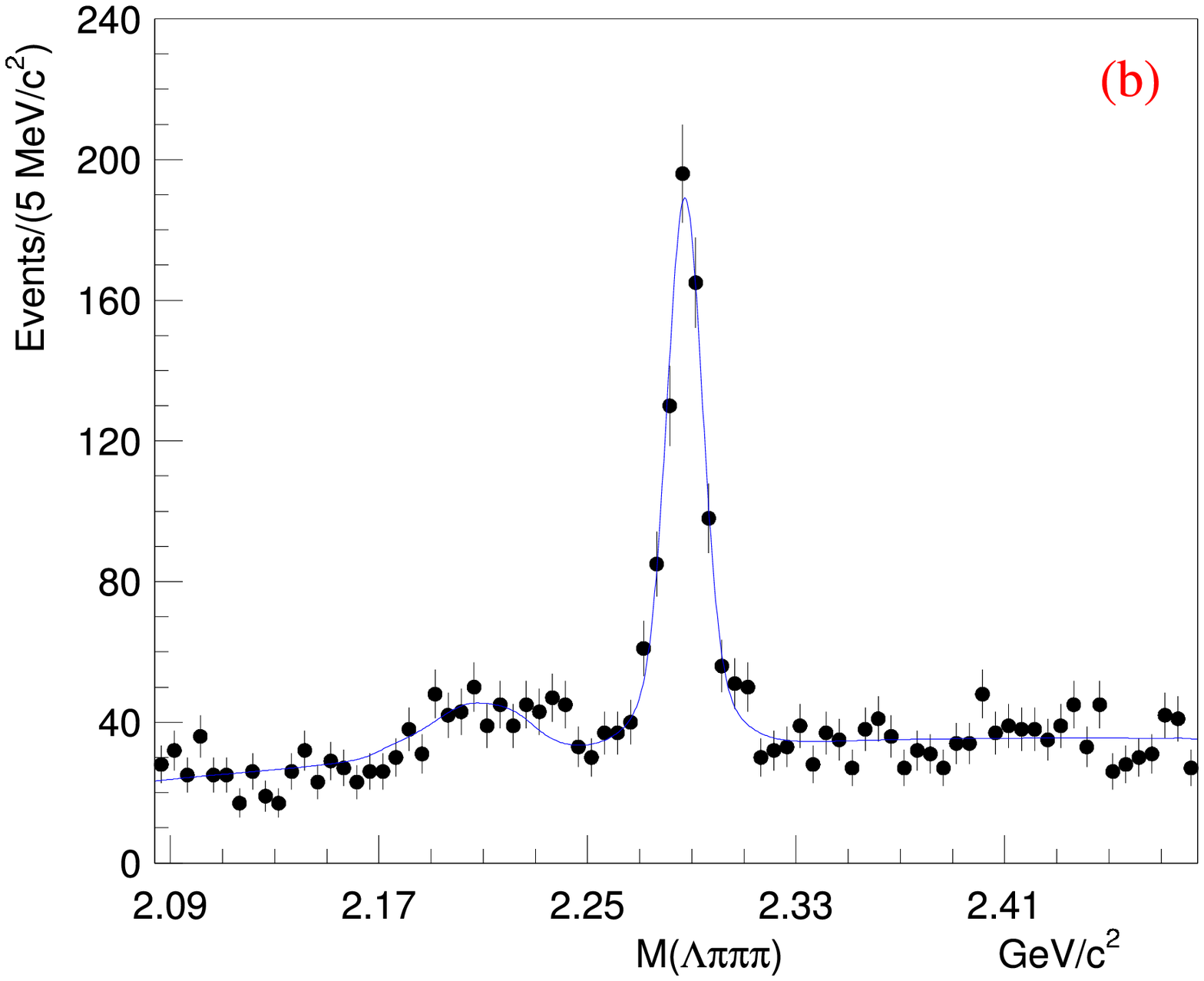}
     \vspace{-0.7cm}
  \caption{Invariant mass distributions for (a) $\Lambda^+_c \rightarrow 
  \Lambda \pi^+ \pi^+ \pi^-$ and (b) 
  $\Lambda^+_c \rightarrow \Lambda \pi^+ \pi^+ \pi^-$ for the subresonant
  analysis. The fits are
  described in the text. }
  \label{figtwo:mass}
\end{figure}

We measure the branching ratio of $\Lambda^+_c \rightarrow \Lambda \pi^+ \pi^+
\pi^-$ relative to $\Lambda^+_c \rightarrow p K^- \pi^+$. In Fig.
\ref{figtwo:mass} (a)  the 
$\Lambda \pi^+ \pi^+ \pi^-$
invariant mass distribution for an $L/\sigma_L > 5$ cut is presented.
The confidence level for the decay vertex must be greater than 5\%. 
We also apply a $\cos \theta > -0.9$ cut, where $\theta$
is the angle between the $\Lambda$ momentum in the $\Lambda^+_c$ rest frame 
and the $\Lambda^+_c$ laboratory momentum.

We also note in this decay mode a broad structure around 2.2 GeV/$c^2$ coming 
from the decay mode $\Lambda^+_c \rightarrow \Sigma^0 (\Lambda \gamma) 
\pi^+\pi^+ \pi^-$ where the photon from the $\Sigma^0$ decay has not been
reconstructed. This has been accounted for as in the $\Lambda^+_c \rightarrow
\Lambda \pi^+$ decay.
The components of the fitting function are the same as in the $\Lambda^+_c \rightarrow
\Lambda \pi^+$ case.
The resultant $\Lambda^+_c \rightarrow \Lambda \pi^+ \pi^+ \pi^-$ yield is $1356 \pm 60 $ events. Correcting for the relative
efficiencies estimated by our Monte Carlo simulation, 
we determine the branching ratio to be

\begin{equation} 
\frac{\Gamma(\Lambda^+_c \rightarrow \Lambda \pi^+ \pi^+ \pi^-)}
{\Gamma(\Lambda^+_c \rightarrow p K^-\pi^+)}= 0.508 \pm 0.024~(\mbox{stat.}). 
\end{equation} 

The number of fitted $\Lambda^+_c \rightarrow \Sigma^0 \pi^+ \pi^+ \pi^-$ 
reflection events is 480 $\pm$ 110. Correcting for the relative efficiencies, 
we extract the relative branching ratio:

\begin{equation}
\frac{\Gamma (\Lambda^+_c \rightarrow \Sigma^0 \pi^+ \pi^+ \pi^-)}
{\Gamma (\Lambda^+_c \rightarrow \Lambda \pi^+ \pi^+ \pi^-)} = 0.26 \pm 
0.06~(\mbox{stat.}).
\end{equation}

We have studied the subresonant structure in the decay mode $\Lambda^+_c
\rightarrow \Lambda \pi^+ \pi^+ \pi^-$. Considering our limited statistics,
which would make a coherent analysis difficult, we use an incoherent binned fit 
method~\cite{incoFra} developed by the E687 Collaboration, which assumes the 
final state is an incoherent superposition of subresonant decay modes.

For the resonant substructure analysis of $\Lambda^+_c \rightarrow \Lambda
\pi^+ \pi^+ \pi^-$ we enhance the signal to noise ratio applying an 
$L/\sigma_L > 8$ cut and requiring $1.11 < M(\Lambda)  <1.119$ GeV/$c^2$. 
In Fig. \ref{figtwo:mass} (b) the $\Lambda \pi^+ \pi^+ \pi^-$
invariant mass distribution for events which satisfy these cuts 
is presented. The resultant yield is $594 \pm 31 $ events.

\begin{figure}[!tbp]
     \includegraphics[width=4.55cm,height=6.8cm]{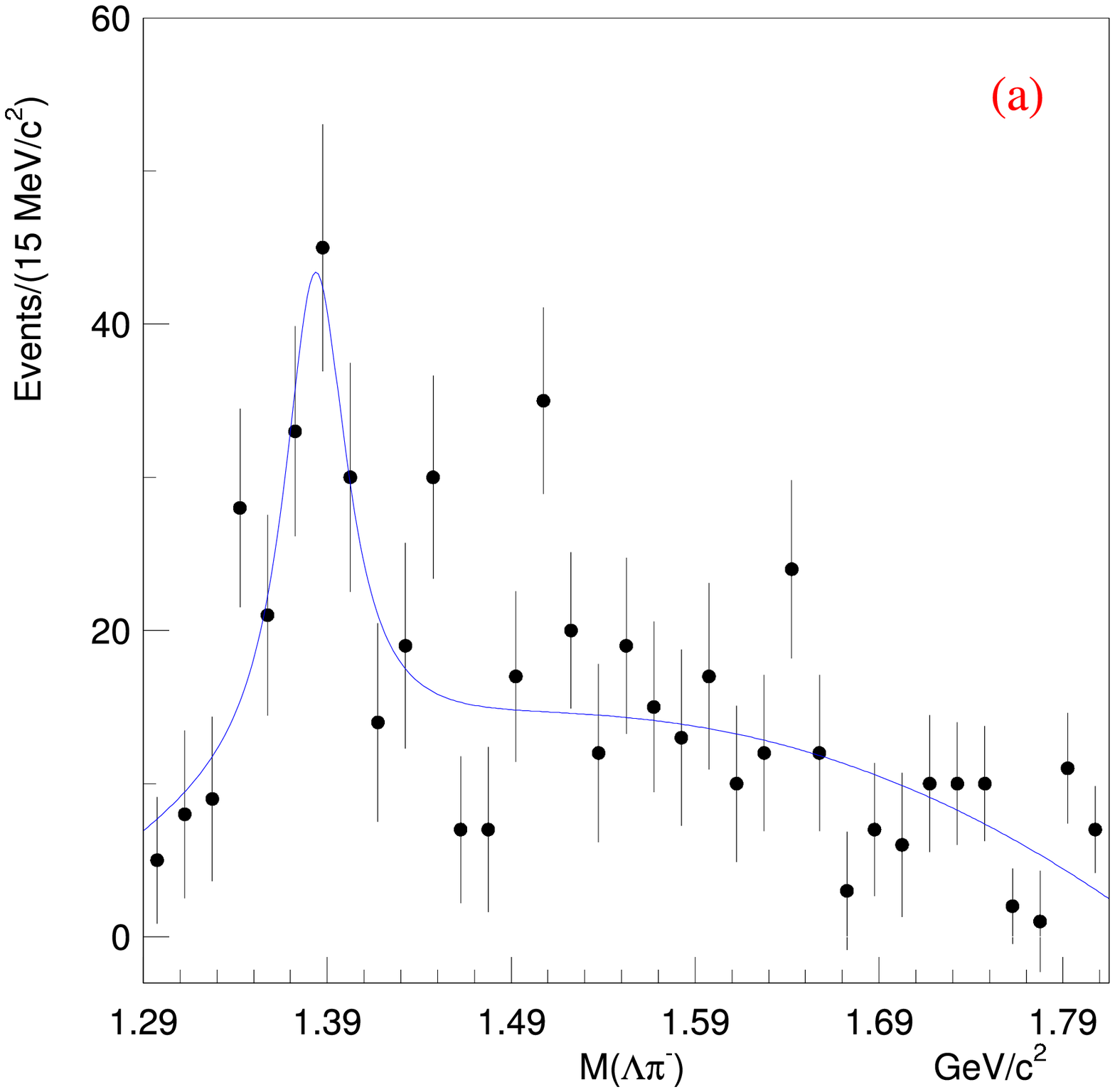}
     \includegraphics[width=4.55cm,height=6.8cm]{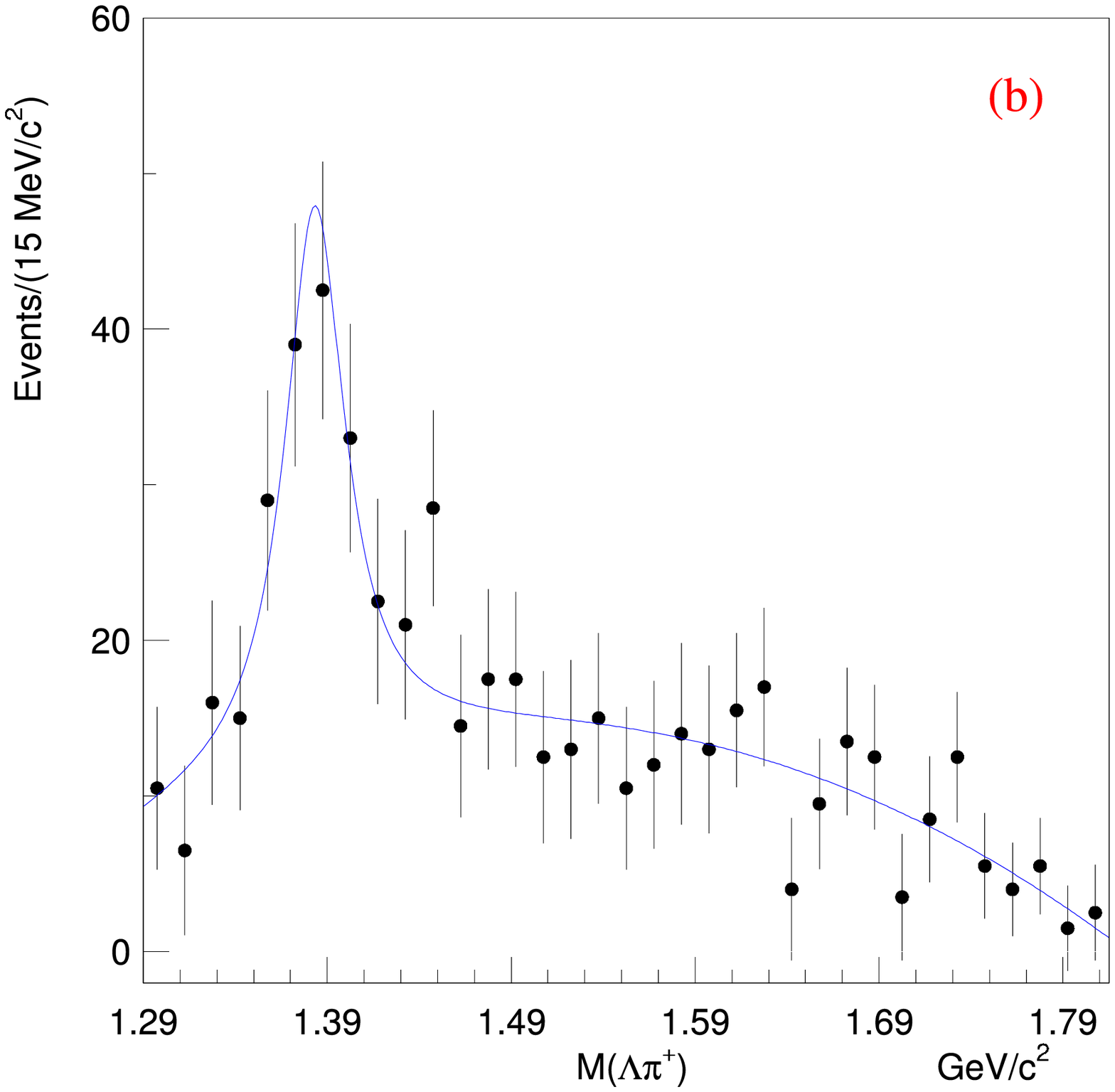}
     \vspace{-0.7cm}
     \includegraphics[width=4.55cm,height=6.78cm]{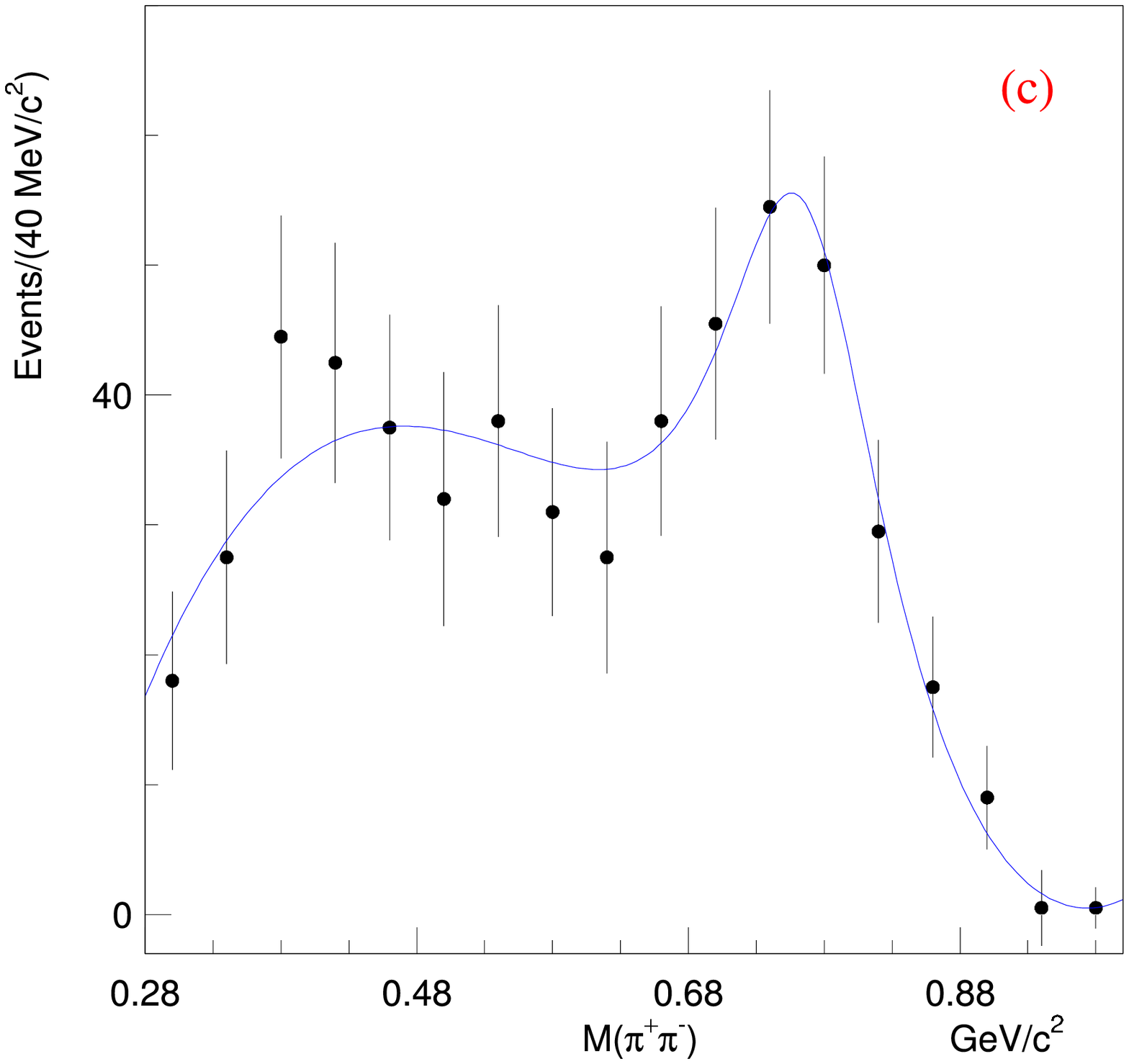}
  \caption{(a) $\Lambda \pi^-$, (b) 
  $\Lambda \pi^+$  and (c) $\pi^+ \pi^-$ invariant mass distributions 
  in the decay mode $\Lambda^+_c \rightarrow \Lambda \pi^+
  \pi^+ \pi^-$. The yields are, respectively, 143 $\pm$ 27, 149 $\pm$ 28 and 317 $\pm$ 68. }
  \label{figrisoP}
\end{figure}

A study of the two-body invariant mass distributions was done to better identify
which resonances may contribute to the $\Lambda \pi^+ \pi^+ \pi^-$ decay channel.
In Fig. \ref{figrisoP} the two body $\Lambda \pi^-$, $\Lambda \pi^+$
and $\pi^+ \pi^-$ invariant mass distributions provide evidence for the $\Sigma(1385)^{\pm}$ and $\rho(770)^0$
resonances. For this study we require the $\Lambda \pi^+ \pi^+ \pi^-$ invariant
mass to be within 2$\sigma$ (18 MeV/$c^2$) of the $\Lambda^+_c$ nominal mass and we perform a sideband 
subtraction to reduce the background. The fits are performed using 
Breit-Wigners for the signal shape, with the mean and width fixed to the Monte Carlo values, 
and Chebychev polynomials for the backgrounds. 

For subresonant modes in the resonant analysis we therefore consider the
channels $\Sigma(1385)^- \pi^+ \pi^+$, $\Sigma(1385)^+ \pi^+ \pi^-$,
$\Lambda \rho(770)^0 \pi^+$ and $\Sigma(1385)^+ \rho(770)^0$, plus a
nonresonant channel  $(\Lambda \pi^+ \pi^+ \pi^-)_{NR}$. All states not
explicitly considered are assumed to be included in the nonresonant channel.

We determine the acceptance corrected yield into each subresonant mode using a
weighting technique whereby each event is weighted by its kinematic values in
the three submasses  ($\Lambda \pi^-$), ($\Lambda \pi^+$) and ($\pi^+ \pi^-$).
We construct eight population bins depending on whether each of the three
submasses falls into the expected resonance peak (within the nominal
width). From a Monte Carlo simulation of
each subresonant mode $\alpha$, we compute the bin population $n_i$ in the 
eight bins and we calculate a transport matrix $T_{i\alpha}$ between 
the number of generated Monte Carlo events $Y_{\alpha}$ and the bin populations:

\begin{equation}
n_i=\sum_{\alpha} T_{i\alpha} Y_{\alpha}.
\end{equation}

The elements of the $T$ matrix can be summed to give the efficiency 
$\epsilon_{\alpha}$ for each mode:

\begin{equation}
\centering
\epsilon_{\alpha}=\sum_i T_{i\alpha}.
\end{equation}  

This Monte Carlo determined matrix is inverted to create a new weighting matrix
which multiplies the bin populations to produce efficiency corrected yields. 
Each data event can then be weighted according to its values in the submass
bins. Once the weighted distributions for each of the five modes have been
generated, we determine the acceptance corrected yields by fitting the
distributions with two Gaussians with the same mean and a second order Chebychev
polynomial for the background. Using incoherent Monte Carlo mixtures of the five
subresonant modes we verify that the method is able to correctly reproduce the
generated mixtures of the different modes. 

\begin{figure}[!h]
\begin{center}
\includegraphics[width=4.55cm,height=6.3cm]{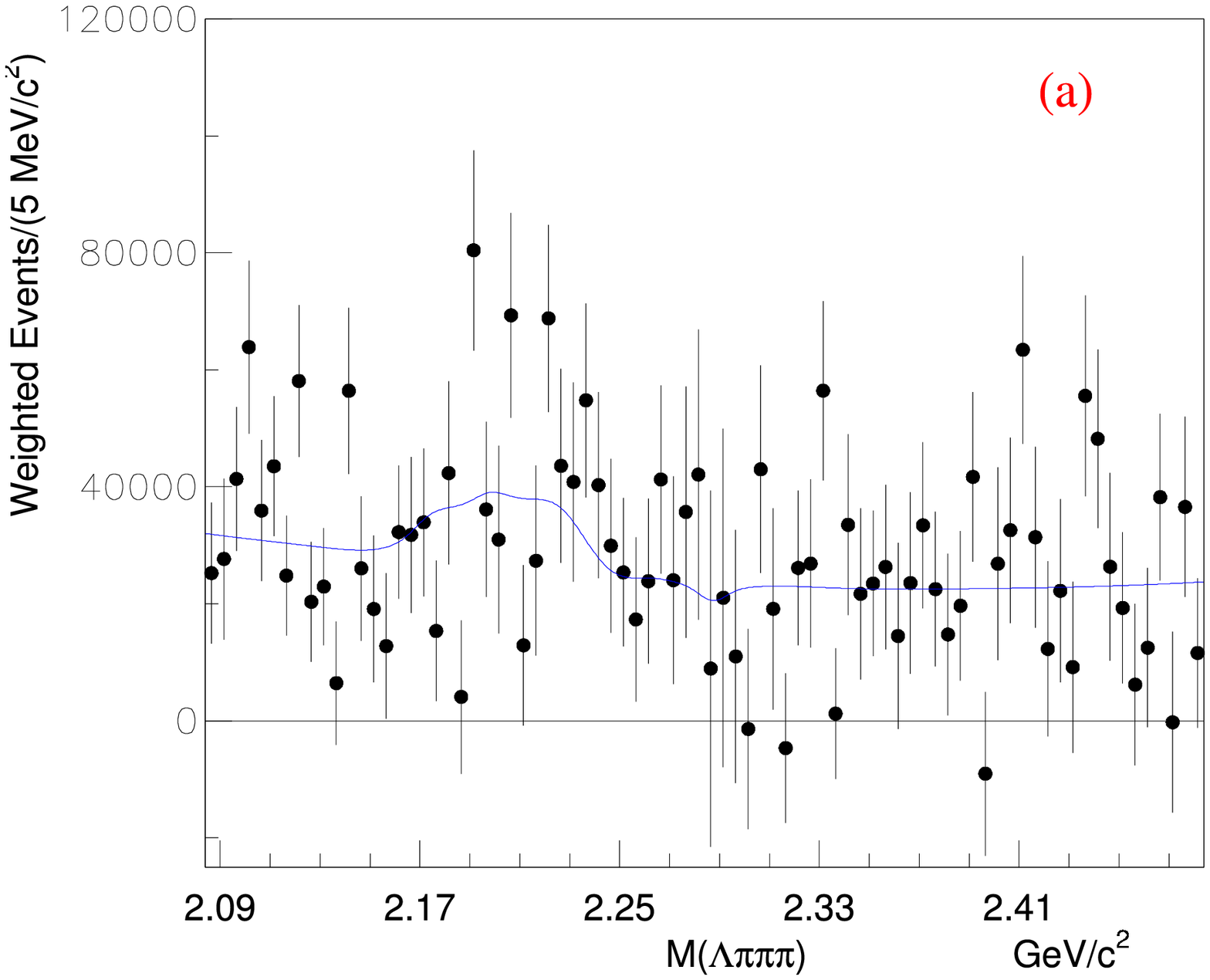}
\includegraphics[width=4.55cm,height=6.3cm]{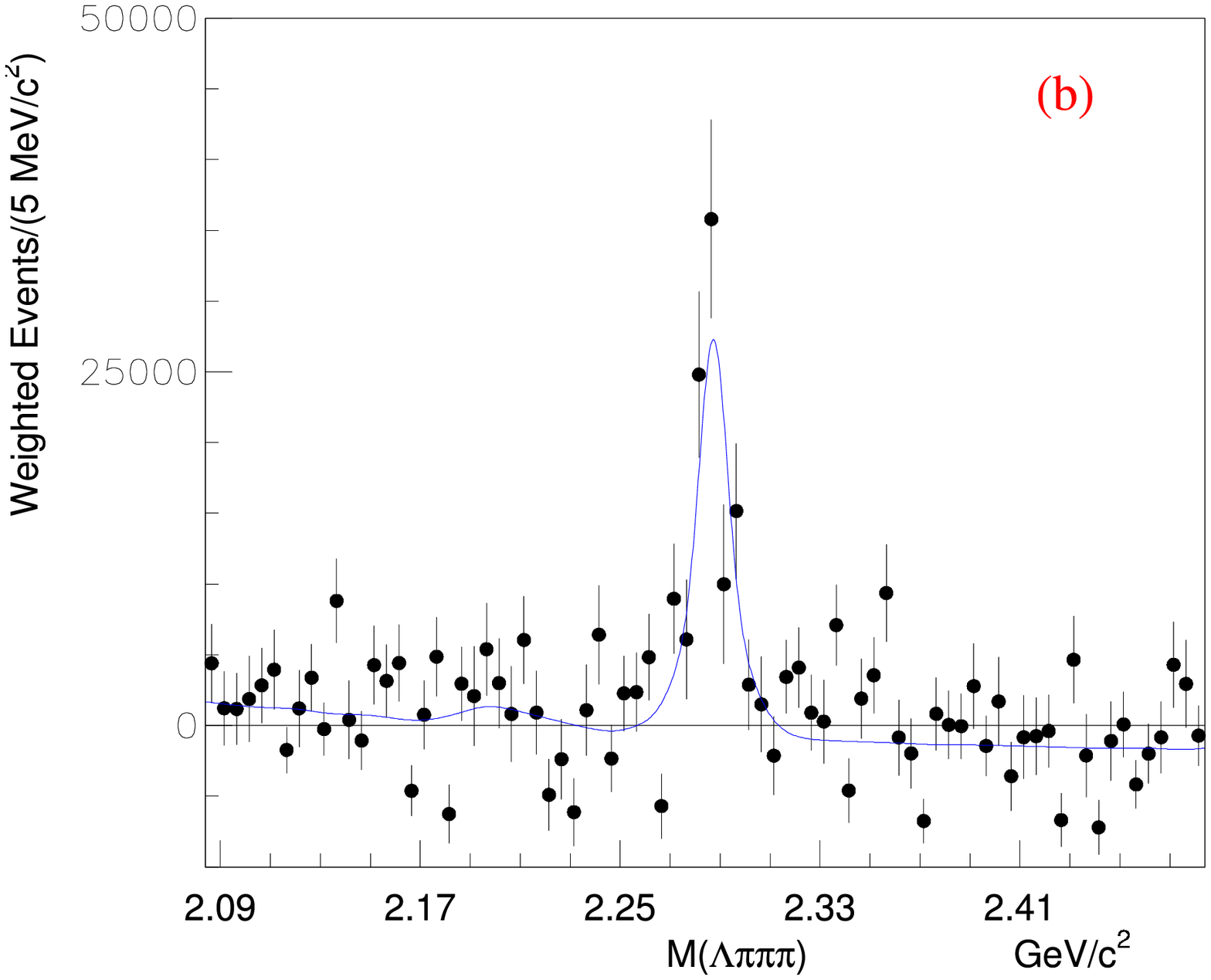}
\includegraphics[width=4.55cm, height=6.27cm]{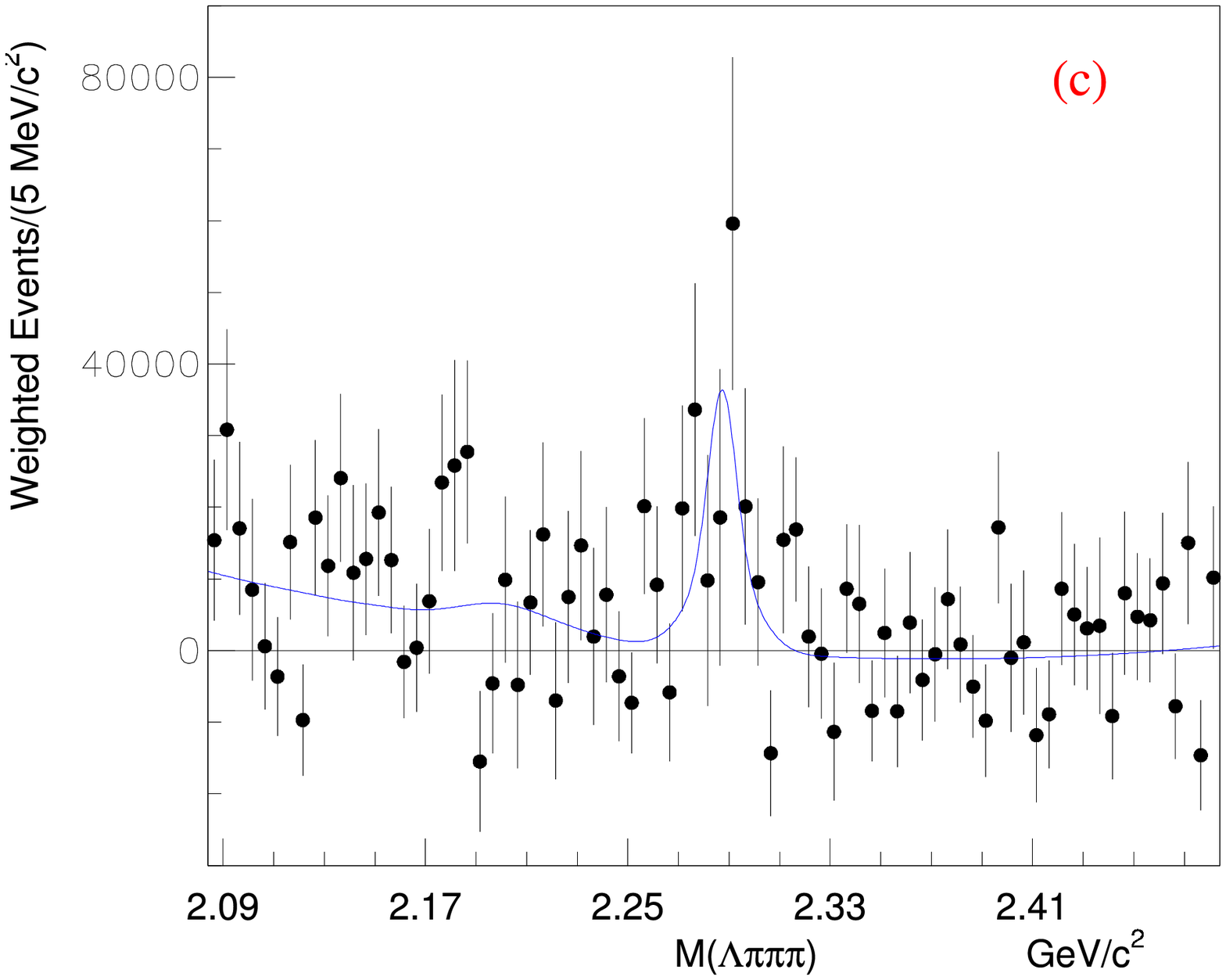}
\includegraphics[width=4.55cm, height=6.3cm]{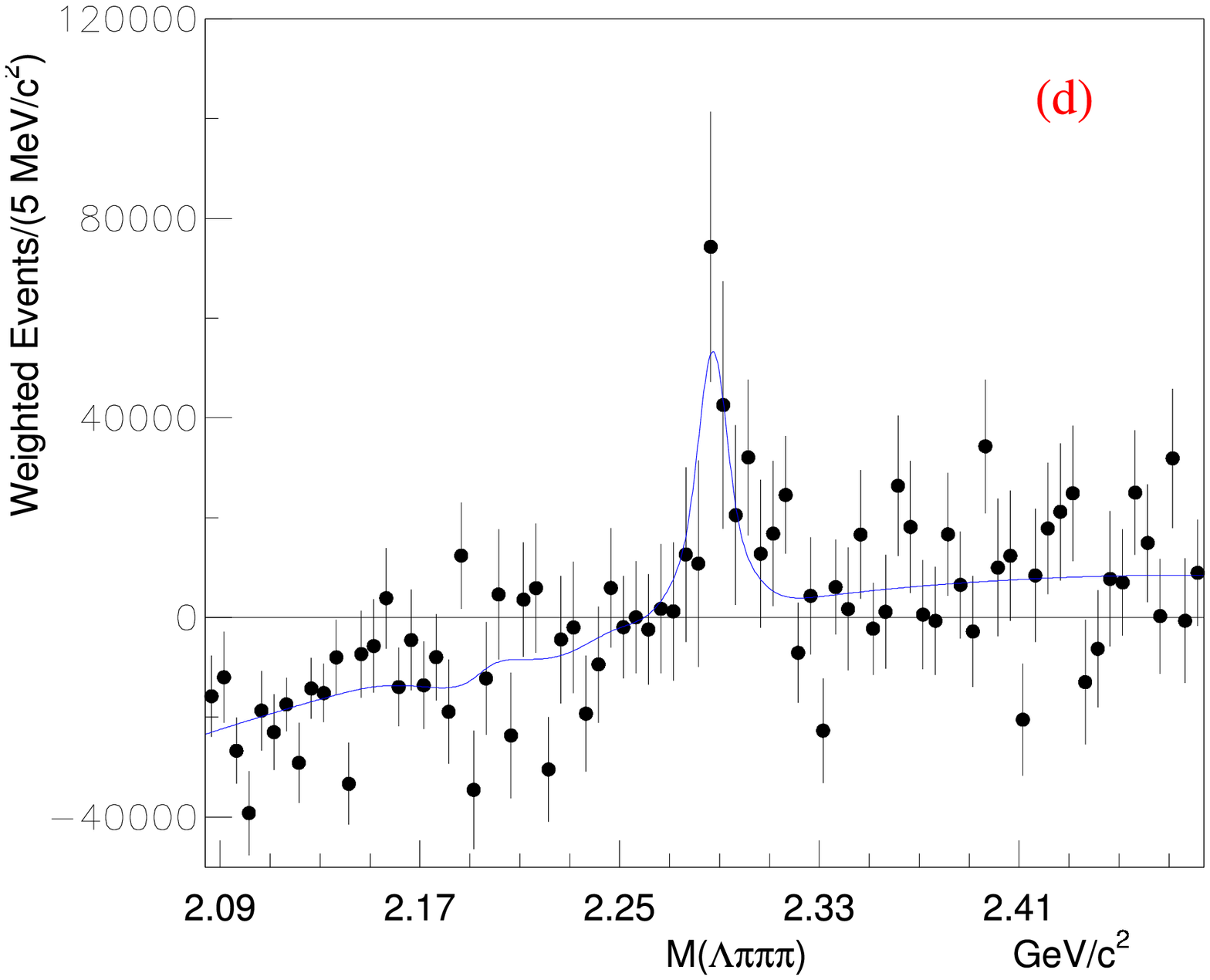}
\includegraphics[width=4.55cm, height=6.3cm]{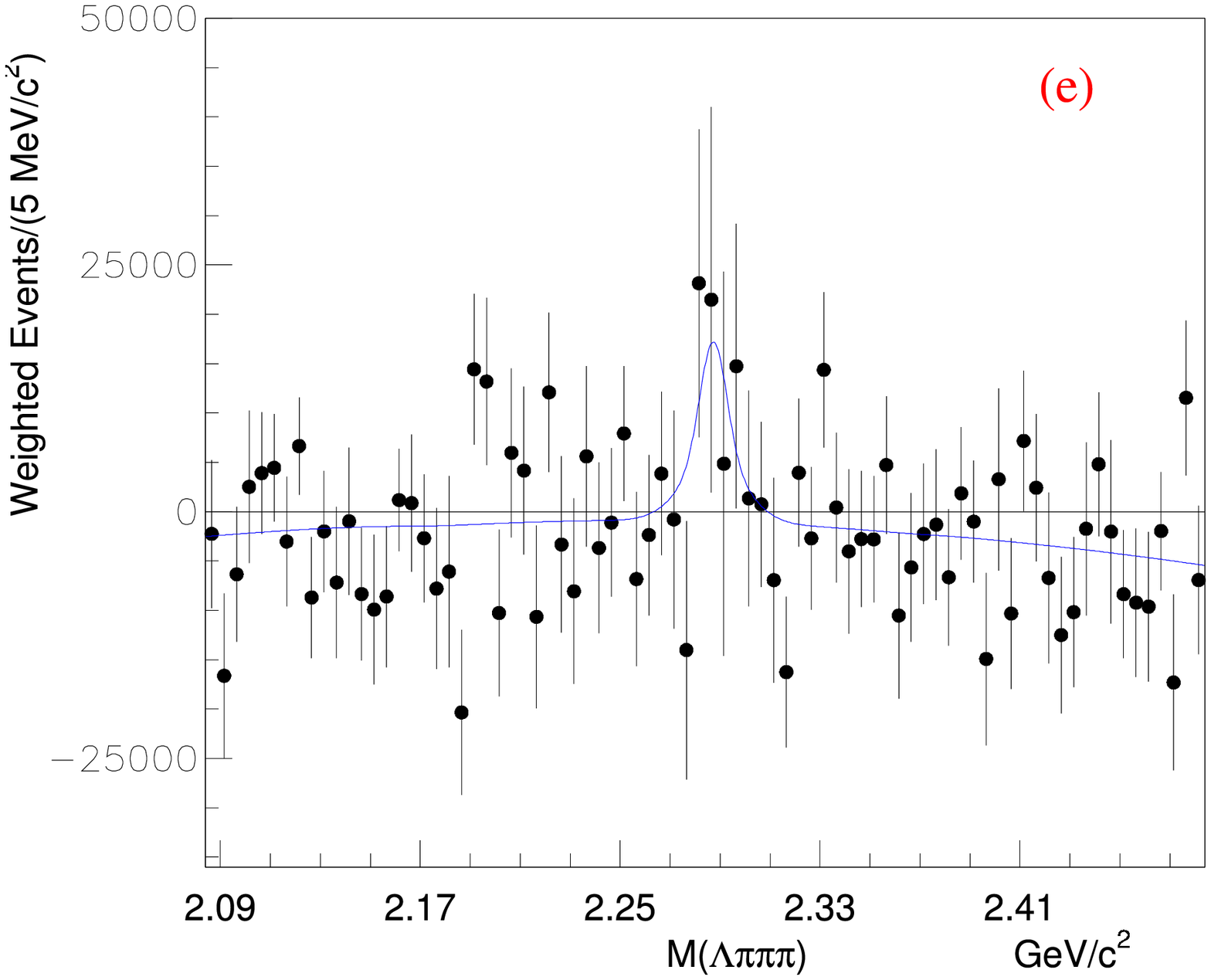}
\includegraphics[width=4.55cm, height=6.3cm]{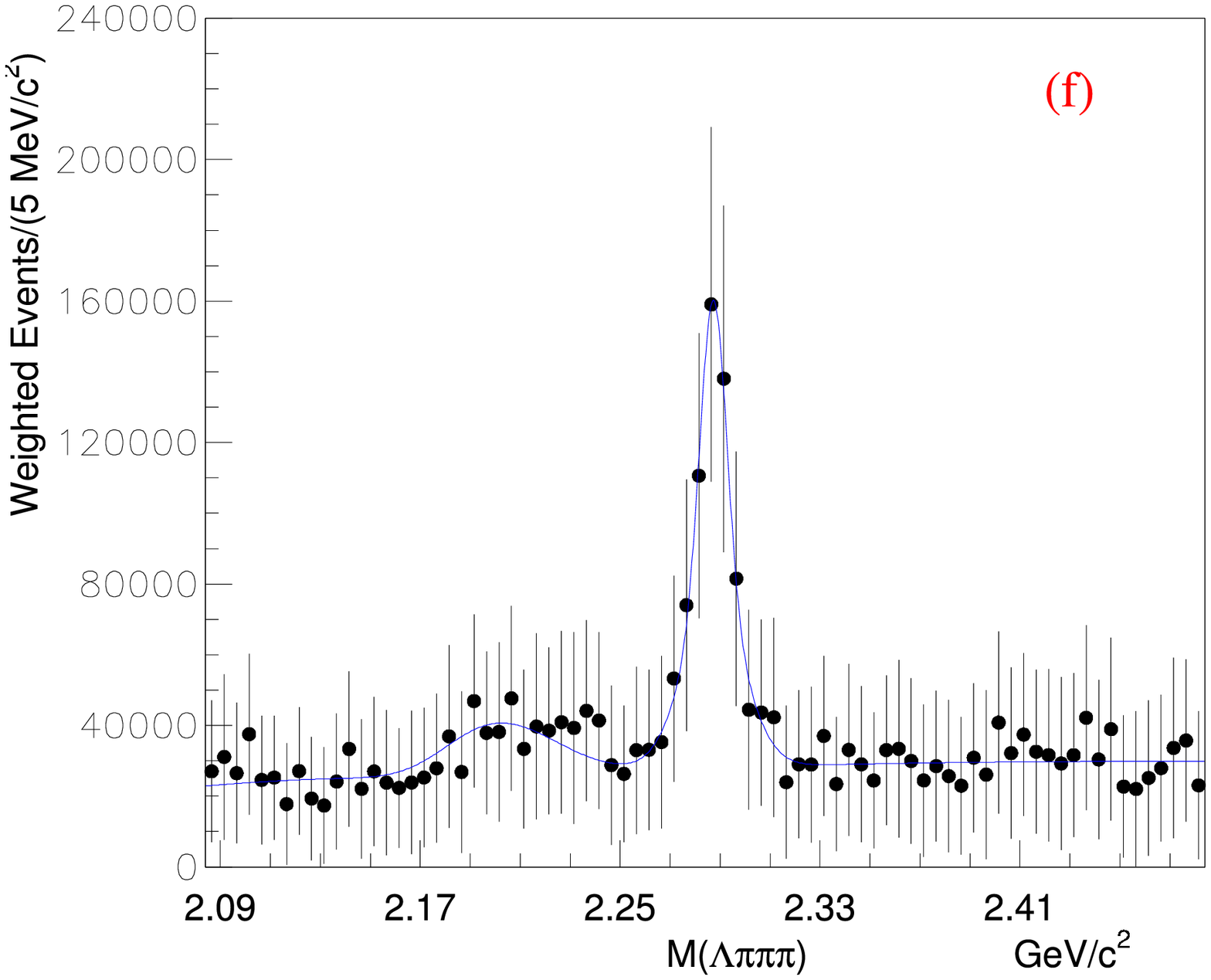}
\end{center}
     \vspace{-0.7cm}
  \caption{$\Lambda^+_c \rightarrow \Lambda \pi^+ \pi^+ \pi^-$ weighted 
  invariant mass distributions for (a) $(\Lambda \pi^+ \pi^+ \pi^-)_{NR}$, 
  (b) $\Sigma(1385)^- \pi^+ \pi^+$, (c) $\Sigma(1385)^+ \pi^+ \pi^-$, (d) 
$\Lambda \rho(770)^0 \pi^+$, (e) $\Sigma(1385)^+ \rho(770)^0$, (f) inclusive sum
of all five modes.}
\label{figtwo:tag}
\end{figure}

The results for the $\Lambda \pi^+ \pi^+ \pi^-$ decay are summarized in Table
\ref{anariso}. The five weighted histograms are shown in Fig. \ref{figtwo:tag},
where Fig. \ref{figtwo:tag} (f) is the weighted distribution for the sum of all subresonant
modes. The systematic uncertainty for the subresonant fractions is estimated
varying the width of the resonance peaks in the construction of the kinematic
bins.    
The goodness of fit is evaluated by calculating a $\chi^2$ for the hypothesis of
consistency between the model predictions and the observed data yields in each
of the 8 submass bins. We obtain a $\chi^2$ of 7.86 (for 3 degrees of freedom) 
and a confidence level of about 5\%. 

\begin{table}[tbp]
\centering
\caption{Fractions relative to the inclusive mode for the subresonant structure
of the $\Lambda^+_c \rightarrow \Lambda \pi^+ \pi^+ \pi^-$ decay mode.}
\vspace{0.2cm}
\begin{tabular}{|c|c|}
\hline
\hline
Subresonant Mode &  Fraction of $\Lambda^+_c \rightarrow \Lambda \pi^+ \pi^+
\pi^-$ \\
\hline
$(\Lambda \pi^+ \pi^+ \pi^-)_{NR} $ &   $<$ 0.30 @90\% CL\\
$\Sigma^{*-} \pi^+ \pi^+$   &  0.21 $\pm$  0.03 $\pm$ 0.02\\
$\Sigma^{*+} \pi^+ \pi^-$  &  0.28 $\pm$ 0.10 $\pm$ 0.08\\
$\Lambda \pi^+ \rho$ &  0.40 $\pm$  0.12 $\pm$ 0.12\\
$\Sigma^{*+} \rho$ &  0.14 $\pm$ 0.09 $\pm$ 0.07\\
\hline
\hline
\end{tabular}
\label{anariso}
\end{table}

\section{\label{sec:level51} The $\Lambda^+_c \rightarrow 
\Lambda \overline{K}\,\!^0 K^+$ decay mode}

\begin{figure}[!ht]
  \begin{center}
     \includegraphics[width=6.825cm,height=7.05cm]{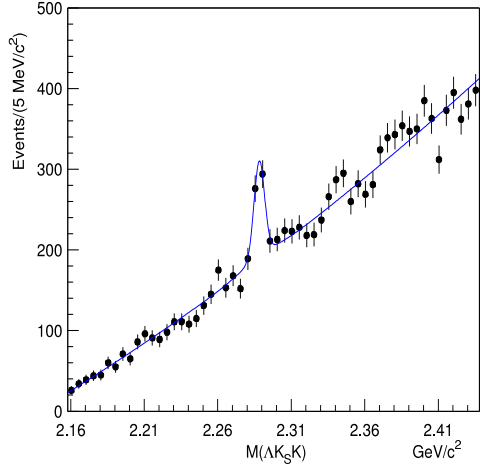}
     \vspace{-0.7cm}
  \caption{Invariant mass distribution for $\Lambda^+_c \rightarrow 
  \Lambda K^0_S K^+$. The fit is
  described in the text. }
  \label{figthree:mass}
\end{center}
\end{figure}

We measure the branching ratio of $\Lambda^+_c \rightarrow \Lambda  
\overline{K}\,\!^0 K^+$
relative to $\Lambda^+_c \rightarrow p K^- \pi^+$. The $\overline{K}\,\!^0$ are detected through
 $K^0_S$'s. Due to the limited phase
space, the signal can be observed without the need for a $L/\sigma_L$ or decay
vertex confidence level cut. In Fig. \ref{figthree:mass}
the $\Lambda K^0_S K^+$ invariant mass distribution is presented.

The fit is performed using two Gaussians with the same mean for the signal 
and a second order Chebychev polynomial for the background. 
The ratio of yields and 
the resolutions of the two Gaussians are fixed to the Monte Carlo values.
The resultant yield is $251 \pm 31 $ events. Correcting for the relative
efficiencies estimated by our Monte Carlo simulation, 
we determine the branching ratio to be

\begin{equation} 
\frac{\Gamma(\Lambda^+_c \rightarrow \Lambda \overline{K}\,\!^0 K^+)}
{\Gamma(\Lambda^+_c \rightarrow p K^-\pi^+)}= 0.142 \pm 0.018~(\mbox{stat.}). 
\end{equation} 

The Belle collaboration~\cite{Belle} has recently shown evidence of the resonant contribution 
$\Lambda^+_c \rightarrow 
\Xi(1690)^0 K^+$ in the decay $\Lambda^+_c \rightarrow 
\Lambda K^0_S K^+$ with the $\Xi(1690)^0$ reconstructed in $\Lambda K^0_S$. In our analysis, the $\Xi(1690)^0 K^+$
events are selected using the same cuts used for the $\Lambda K^0_S K^+$ mode; 
the $\Lambda K^0_S K^+
$ invariant mass is required to be within 2$\sigma$ (10 MeV/$c^2$) of the $\Lambda^+_c$ nominal
mass. A sideband subtraction is performed to reduce the combinatoric background
under the $\Lambda^+_c$ signal region.

\begin{figure}
  \begin{center}
     \includegraphics[width=6.825cm,height=7.05cm]{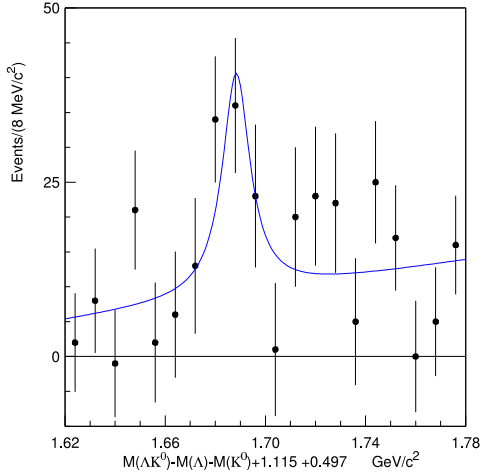}
     \vspace{-0.7cm}
  \caption{Invariant mass distribution for $\Lambda K^0_S$ in the decay $\Lambda^+_c \rightarrow 
  \Lambda K^0_S K^+$. The fit is
  described in the text.}
  \label{figfour:mass}
\end{center}
\end{figure}

The $\Lambda K^0_S$ invariant mass distribution is shown in Fig. \ref{figfour:mass}.
The fit is performed using a Breit-Wigner function for the signal  
and a first order Chebychev polynomial for the background. 
The mean and the width of the Breit-Wigner are fixed to the Monte Carlo
values.\footnote{The $\Xi(1690)^0$ is generated in our Monte Carlo simulation with
a mass of 1.688 GeV/$c^2$ and a width of 10 MeV/$c^2$.}
The resultant yield is $84 \pm 24$ events. 

We measure the branching ratio relative to $\Lambda^+_c \rightarrow \Lambda \overline{K}\,\!^0 K^+$ 
to be

\begin{equation} 
\frac{\Gamma(\Lambda^+_c \rightarrow \Xi(1690)^0 K^+)}
{\Gamma(\Lambda^+_c \rightarrow \Lambda \overline{K}\,\!^0 K^+)} \times 
B(\Xi(1690)^0 \rightarrow \Lambda \overline{K}\,\!^0)
= 0.33 \pm 0.10~(\mbox{stat.}). 
\end{equation} 

\section{\label{sec:level61} Systematic studies}

The systematic effects are evaluated after investigation of different
sources: uncertainties in the reconstruction efficiency and in the resonant
substructure for multibody decays and the choice of fitting conditions. 

To
determine the systematic error due to the reconstruction efficiency we follow a
procedure based on the S-factor method used by the Particle Data Group~\cite{PDG}. For
each mode we split the data sample into independent subsamples based on
$\Lambda^+_c$ momentum, data-taking period, particle-antiparticle, 
 significance of separation between production and decay vertices and different
 $\Lambda$ and $K^0_S$ categories, based on the location and geometry of the
 neutral particle decay. These splits provide a check on the Monte Carlo
 simulation of charm production, of the vertex detector and of different
 variables employed in the event selection. We define the split sample variance
 as the difference between the scaled variance and the statistical variance if
 the former exceeds the latter. The method is described in detail in 
 Reference~\cite{split}.

 Considering the large uncertainty on the measured subresonant fractions in the 
 multibody decays, we also vary these fractions in the Monte Carlo simulation 
 and we use the variance in the branching ratios as a contribution to the systematic error.   

We measure the systematic uncertainty due to fitting conditions using a fit
variation technique, which includes variations in bin size, fitting range,
background and signal shapes (different order of the Chebychev polynomial,
leaving the two Gaussian parameters free in the fit or using a single 
Gaussian for the signal). 

We also include a systematic error contribution from 
the absolute tracking efficiency for the different multiplicities in the final
states. In Table \ref{systtable} we summarize the systematic 
uncertainty for each mode. Several measurements for the modes reported here are  present in the
literature~\cite{Lpi1,Lpi2,Lpi3,Lpi4,L3pi1,Spi1,cleolksk}. In Table \ref{brtable} we present the FOCUS 
results with a comparison to the PDG values~\cite{PDG}.  
  
\begin{table}[tbp]
\centering
\caption{The systematic uncertainties from the Monte Carlo simulation, the
fitting condition and the the total for each mode.}
\vspace{0.2cm}
\begin{tabular}{|c|c|c|c|c|c|c|}
\hline
\hline
Mode &  Simulation & Subresonances & Tracking & Fit & Total\\
\hline
$\frac{\Gamma(\Lambda^+_c \rightarrow \Lambda \pi^+)}{\Gamma
(\Lambda^+_c \rightarrow p K^- \pi^+)}$ &  0.017  & --- & 0.005 & 0.008 & 0.020 \\
$\frac{\Gamma(\Lambda^+_c \rightarrow \Sigma^0 \pi^+)}{\Gamma 
(\Lambda^+_c \rightarrow \Lambda \pi^+)}$ &  0.19 & --- & --- & 0.04 & 0.19 \\
$\frac{\Gamma(\Lambda^+_c \rightarrow \Lambda \pi^+ \pi^+ \pi^-)}
{\Gamma(\Lambda^+_c \rightarrow
p K^- \pi^+)}$ &  0.016 & 0.010 & --- & 0.014 & 0.024\\
$\frac{\Gamma(\Lambda^+_c \rightarrow \Sigma^0 \pi^+\pi^+ \pi^-)}
{\Gamma(\Lambda^+_c \rightarrow \Lambda \pi^+ \pi^+ \pi^-)}$ &  
0.08 & --- & --- & 0.03 & 0.09 \\
$\frac{\Gamma(\Lambda^+_c \rightarrow \Lambda \overline{K}\,\!^0 K^+)}
{\Gamma(\Lambda^+_c \rightarrow
p K^- \pi^+)}$ &  0.021 & 0.001 & 0.004 & 0.005 & 0.022 \\
$\frac{\Gamma(\Lambda^+_c \rightarrow \Xi(1690)^0(\Lambda \overline{K}\,\!^0) K^+)}
{\Gamma(\Lambda^+_c \rightarrow \Lambda \overline{K}\,\!^0 K^+)}$ & --- & --- & --- & 
0.04 & 0.04\\
\hline
\hline
\end{tabular}
\label{systtable}
\end{table}

\section{Conclusions}

We have investigated and measured the branching ratios of several $\Lambda^+_c$ 
Cabibbo-favored decay modes containing the $\Lambda$ hyperon in the final state.
These modes are $\Lambda^+_c \rightarrow \Lambda 
\pi^+, \Lambda^+_c \rightarrow \Lambda \pi^+\pi^+\pi^-$ and $
\Lambda^+_c \rightarrow \Lambda \overline{K}\,\!^0 K^+$. From the fit to the 
first two modes, we are also able to extract the relative branching ratios 
of the two decays $\Lambda^+_c \rightarrow \Sigma^0 \pi^+$ and $\Lambda^+_c \rightarrow \Sigma^0
\pi^+\pi^+\pi^-$. These measurements are an improvement over previous results 
for the same decay modes. 
We report a new measurement of the subresonant mode 
$\Lambda^+_c \rightarrow \Xi(1690)^0 K^+$ consistent
with the recent Belle result. We have also performed an analysis of the 
subresonant structure of the decay
$\Lambda^+_c \rightarrow \Lambda \pi^+ \pi^+ \pi^-$. 
We observe a small nonresonant component and the
presence of vector resonances in the dominant modes, as it has been observed in
most charm meson decays.

\section{\label{sec:level8}Acknowledgments} 
~
We wish to acknowledge the assistance of the staffs of Fermi National
Accelerator Laboratory, the INFN of Italy, and the physics departments
of
the
collaborating institutions. This research was supported in part by the
U.~S.
National Science Foundation, the U.~S. Department of Energy, the Italian
Istituto Nazionale di Fisica Nucleare and 
Ministero dell'Istruzione, dell'Universit\`a e della 
Ricerca, the Brazilian Conselho Nacional de
Desenvolvimento Cient\'{\i}fico e Tecnol\'ogico, CONACyT-M\'exico, and
the Korea Research Foundation of the  
Korean Ministry of Education.

\begin{sidewaystable}
\hspace{-0.5cm}
\centering
\caption{FOCUS results compared to previous measurements. No direct measurement
exists for the relative branching ratios 
$\Gamma(\Lambda^+_c \rightarrow \Sigma^0 \pi^+)/\Gamma(\Lambda^+_c \rightarrow
\Lambda \pi^+)$ and $\Gamma(\Lambda^+_c \rightarrow \Sigma^0 \pi^+ \pi^+ \pi^-)/\Gamma(\Lambda^+_c 
\rightarrow \Lambda \pi^+ \pi^+ \pi^-)$. The relative efficiency includes the 
branching fractions into the observed final state particles.}
\vspace{0.25cm}
\begin{tabular}{|ccccccc|}
\hline
\hline
\vspace{-0.2cm}
$\Lambda^+_c$ Decay & Signal & $\Lambda^+_c$ Reference & Reference & Relative
 & FOCUS  & PDG~\cite{PDG}\\
Mode & Yield & Mode & Yield & Efficiency & & \\
\hline
$\Lambda \pi^+$ & 750 $\pm$
 44 & $
p K^- \pi^+$ & 16447 $\pm$ 193 & 0.209 $\pm$ 0.001 & 0.217 $\pm$ 0.013 $\pm$ 0.020 & 0.180 $\pm$ 0.032 \\
$\Sigma^0 \pi^+$ & 919 $\pm$ 92 &
$\Lambda \pi^+$ & 750 $\pm$ 44 & 1.119 $\pm$ 0.001 & 1.09 $\pm$ 0.11 $\pm$ 0.19 & 1.11 $\pm$ 0.49\\
$\Lambda \pi^+ \pi^+ \pi^-$ & 1356 $\pm$ 60 &
$p K^- \pi^+$ & 12898 $\pm$ 147 & 0.207 $\pm$ 0.001 & 0.508 $\pm$ 0.024 $\pm$ 0.024 &  0.66 $\pm$ 0.11\\
$\Sigma^0 \pi^+\pi^+ \pi^-$ & 480 $\pm$ 110 &
$\Lambda \pi^+ \pi^+ \pi^-$ & 1356 $\pm$ 60 & 1.375 $\pm$ 0.001 &  
0.26 $\pm$ 0.06 $\pm$ 0.09 & 0.33 $\pm$ 0.16 \\
$\Lambda \overline{K}\,\!^0 K^+$ & 251 $\pm$ 31 &
$p K^- \pi^+$ & 10952 $\pm$ 132 & 0.161 $\pm$ 0.001 & 0.142 $\pm$ 0.018 $\pm$
0.022  & 0.12 $\pm$ 0.02 $\pm$ 0.02\\
$\Xi(1690)^0(\Lambda \overline{K}\,\!^0) K^+$ & 84 $\pm$ 24 &
$\Lambda \overline{K}\,\!^0 K^+$ & 251 $\pm$ 31 & 1.00 
& 0.33 $\pm$ 0.10 $\pm$ 0.04 & 0.26
$\pm$ 0.08 $\pm$ 0.03\\
\hline
\hline
\end{tabular}
\label{brtable}
\end{sidewaystable}

\end{document}